\documentclass[11pt]{article}
\usepackage{amsmath}
\usepackage{amssymb}

\usepackage{calc}
\newcommand{\inst}[1]{\mbox{$^{\text{\textnormal{#1}}}$}}

\begin{document}
\numberwithin{equation}{section}

%%%%%%%%%%%%%%%%%%%%%%%%%%%%%%%%%%
\begin{flushright}
EPHOU-13-008\\
September, 2013
\end{flushright}
\mbox{}\\\bigskip\bigskip

\begin{center}
%% ----- title -----
{\LARGE Off-shell Invariant D=N=2 Twisted Super Yang-Mills Theory with a
 Gauged Central Charge without Constraints }\\[8ex]

%% ----- author -----
{\large
Keisuke Asaka\inst{a}\footnote{\texttt{asaka@high.hokudai.ac.jp}},
Junji Kato\inst{b}\footnote{\texttt{jkato@particle.sci.hokudai.ac.jp}},
Noboru Kawamoto\inst{b}\footnote{\texttt{kawamoto@particle.sci.hokudai.ac.jp}}
{\normalsize and}
Akiko Miyake\inst{c}\footnote{ \texttt{miyake@ippan.kushiro-ct.ac.jp}} }\\[4ex]
          
%% ----- institute -----
{\large\itshape
\inst{a} Institute for the Advancement of Higher Education, Hokkaido
 University\\
Sapporo, 060-0817 Japan \\[3ex]
\inst{b} Department of Physics, Hokkaido University\\
Sapporo, 060-0810 Japan\\[3ex]
\inst{c} Department of General Education, Kushiro National College of
 Technology\\
Kushiro, 084-0916 Japan
}
\end{center}
\bigskip\bigskip

\begin{abstract}
We formulate N=2 twisted super Yang-Mills theory with a gauged central
 charge by superconnection formalism in two dimensions. 
We obtain off-shell invariant
 supermultiplets and actions with and without constraints, which is in
 contrast with the off-shell invariant D=N=4 super Yang-Mills
 formulation with unavoidable constraints. 
\end{abstract}

%PACS codes:

%Keywords:
 
\newpage
\section{Introduction}

Supersymmetry (SUSY) is one of the most important guiding principle in
contemporary particle physics. 
In particular it has been recognized that D=N=4 super Yang-Mills (SYM) theory
plays a crucial role in string motivated gauge theory formulations \cite{Maldacena}.
It has also been recognized that D=N=2 and D=N=4 SYM
formulation play a special role for Dirac-K\"{a}hler twisting
procedure \cite{Dirac Kaehler}, which gives links to quantization and supersymmetry
\cite{Dirac Kaehler,topological twist},
and to lattice supersymmetry \cite{twist}.
There is a long-standing question if one can find a superspace
formulation for D=N=4 SYM to obtain off-shell invariant formulation.
It is known that D=N=4 SYM with SU(4) R-symmetry can be formulated only
on-shell level \cite{only on-shell} while one can find off-shell invariant SYM formulation if
we introduce a central charge and change the R-symmetry from SU(4) to
USp(4) \cite{central charge,SSW}. In this case we, however, need a constraint equation which
can be seen as remnant of an equation of motion of higher dimensions.
The corresponding superspace formulation has been developed in
\cite{Milewski,Saito,Buchbinder}. Harmonic superspace has also been
developed in the similar context \cite{harmonic}.

One may thus ask a question whether or not such a constraint
 is unavoidable for gauged central charge SYM formulation.
In this paper we investigate D=N=2 off-shell invariant twisted SYM with gauged
central charge by suerconnection formalism \cite{Labastida,KKM}. 
This kind of gauged central charge formulation is called vector-tensor multiplet
and has been investigated intensively \cite{VT}. 
It turns out that for A model ansatz we obtain off-shell invariant
formulation with a constraint. 
Although this constraint can be solved in the case of Abelian gauge group, 
it is not possible to solve the constraint at least locally in the 
case of non-Abelian gauge group \cite{solving constraint}.
For B model ansatz we obtain off-shell invariant formulation without any
constraints.

This paper is organized as follows: In Section 2 we first discuss
twisted superalgebra with central charges. In Section 3 we consider
three models for three different supercurvature ansatz.
Then we obtain supermultiplets and actions in Section 4. 
We then summarize the result and give some discussions in the last section.   

\section{D=N=2 Twisted Superalgebra with Central Charge}

In this section we introduce central charge to N=2 superalgebra in two
dimensions and perform Dirac K\"{a}hler twisting.
We concentrate on Euclidean spacetime in this paper according to the twisting 
procedure \cite{Dirac Kaehler}.
 
\subsection{Superalgebra with central charge}

In two dimensional Euclidean spacetime $\gamma$-matrices satisfying
 $\{\gamma^{\mu},\gamma^{\nu}\}=2\delta^{\mu\nu}$ and charge conjugation
 matrix $C$ can be chosen to satisfy \cite{Kugo-Townsent}
\begin{equation}
C\gamma^{\mu T}C^{-1}=\gamma^{\mu}, \hspace{10pt}C^{T}=C.
\end{equation}
Thus we can choose the representation
of these matrices as
\begin{equation}
 \gamma^{1}=\begin{pmatrix}1&0 \\ 0&-1 \end{pmatrix}\,,\hspace*{10pt}
 \gamma^{2}=\begin{pmatrix}0&1 \\ 1&0 \end{pmatrix}\,,\hspace*{10pt}
 C=1,\hspace*{10pt}
 \gamma^{5}\equiv i\gamma^{1}\gamma^{2}=\begin{pmatrix}0&i \\ -i&0 \end{pmatrix}\,.
\end{equation}
Note that $\gamma^{\mu}$ and $\gamma^{5}$ are symmetric and
antisymmetric matrices, respectively. 

The general extended superalgebra is given by
\begin{align}
 \{Q_{\alpha i},Q_{\beta
 j}\}=&2\delta_{ij}\gamma^{\mu}_{\alpha\beta}P_{\mu}\,,\nonumber \\
[Q_{\alpha i},R_{1}]=&iS_{ij}Q_{\alpha j}\,,\nonumber\\
[Q_{\alpha i},R_{2}]=&{S^{\prime}}_{ij}\gamma^{5}_{\,\alpha\beta}Q_{\beta j}\,,\nonumber\\
[Q_{\alpha i},P_{\mu}]=[R_{i},P_{\mu}]=&[P_{\mu},P_{\nu}]=[R_{i},R_{j}]=0\,,
\end{align}
where $Q_{\alpha i}$ is supercharge and $R_{1}$, $R_{2}$ are generators
of R-symmetry. Here we consider D=N=2 case. Majorana condition is given by $Q_{\alpha i}={Q_{\alpha i}}^{\ast}$. 
Jacobi identities w.r.t. $Q$, $Q$, $R_{1}$ and $Q$, $Q$, $R_{2}$ lead,
respectively, $S_{ij}=-S_{ji}$ and $S^{\prime}_{ij}=-S^{\prime}_{ji}$,
 which means $R_{1}$ and $R_{2}$ generate $U(1)$ symmetry.  

We now introduce possible extra terms as follows:
\begin{equation}
\{Q_{\alpha i},Q_{\beta
 j}\}=2\delta_{ij}\gamma^{\mu}_{\alpha\beta}P_{\mu}+2\delta_{\alpha\beta}U_{ij}+2\gamma^{5}_{\alpha\beta}V_{ij}\,,
\end{equation}
where $U_{ij}=U_{ji}$, $V_{ij}=-V_{ji}$ to be consistent with simultaneous replacements of
$\alpha\leftrightarrow\beta$ and $i\leftrightarrow j$. 
$U_{ij}$ and $V_{ij}$ get the following restrictions according to a Jacobi identity w.r.t. $Q$, $Q$, $R_{1}$:
\begin{equation}
U_{ik}S_{kj}-S_{ik}U_{kj}=0\,,\hspace*{10pt}V_{ik}S_{kj}+S_{ik}V_{kj}=0\,.
\end{equation}
We can then solve the constraints up to an over all constant as
\begin{equation}
 U_{ij}\sim\begin{pmatrix}1&0 \\ 0&1 \end{pmatrix}\,,\hspace*{10pt}
 V_{ij}\sim\begin{pmatrix}0&1 \\-1&0 \end{pmatrix}\,.\label{a10}
\end{equation}  
On the other hand Jacobi identity w.r.t. $Q$, $Q$,
$R_{2}$ leads to the relation: 
\begin{equation}
U_{ik}S^{\prime}_{kj}+S^{\prime}_{ik}U_{kj}=0\,,\hspace*{10pt}
V_{ik}S^{\prime}_{kj}-S^{\prime}_{ik}V_{kj}=0\,,
\end{equation}
which can be solved as
\begin{equation}
 U_{ij}\sim\begin{pmatrix}u^{\prime}&u \\ u&-u^{\prime} \end{pmatrix},\hspace*{10pt}
 V_{ij}=0\,,\label{a11}
\end{equation} 
where $u,u^{\prime}$ are real parameters.
The solutions (\ref{a10}) and (\ref{a11}) are not
compatible. 
In other words we cannot keep both of $R_{1}$ and $R_{2}$
symmetries. 
In case we choose $R_{1} (\equiv R)$ symmetry
we obtain the following algebra: 
\begin{align}
\{Q_{\alpha i},Q_{\beta
 j}\}&=2\delta_{ij}\gamma^{\mu}_{\alpha\beta}P_{\mu}+2\delta_{\alpha\beta}\delta_{ij}U_{
0}+2\gamma^{5}_{\alpha\beta}\gamma^{5}_{ij}V_{5}\,,\label{a7}\\
 [Q_{\alpha i},R]&=iS_{ij}Q_{\alpha j}\,,\label{a8}\\
[U_{0},\text{any}]&=[V_{5},\text{any}]=0\,,
\end{align}
where we identify $U_{0}$ and $V_{0}$ as central charges.
On the other hand if we choose $R_{2}$ symmetry, we cannot carry out
Dirac-K\"{a}hler twisting procedure. 
We thus not choose this case. 

\subsection{Twisted Superalgebra}

Dirac-K\"{a}hler twisting procedure includes two steps: 
Expansion of supercharge by complete set of 
$\gamma$-matrices and redefinition of Lorentz rotation generator
\cite{Dirac Kaehler,Saito,KKM}.

We identify the representations of R-symmetry as that of spinor, and
 treat the extended SUSY suffix and spinor suffix of supercharge
$Q_{\alpha i}$ in the same manner.
We thus expand the charge as 
 \begin{equation}
 Q_{\alpha i}=(1s+\gamma^{\mu}s_{\mu}-i\gamma^{5}\tilde{s})_{\alpha i}\,,\label{a0}
\end{equation}
where $s$, $s_{\mu}$ and $\tilde{s}$ are called twisted supercharges. 
Note that these supercharges can be expressed by the original charge as
\begin{equation}
 s=\frac{1}{2}\textrm{tr\,}Q\,,\hspace*{10pt}
s_{\mu}=\frac{1}{2}\textrm{tr\,}\gamma^{\mu}Q\,,\hspace*{10pt}
\tilde{s}=-\frac{1}{2}\textrm{tr\,}\gamma^{5}Q\,.\label{b1}
\end{equation}
The charges may be looked strange because $s_{\mu}$ has, for example, vector
suffix although it is fermionic charge. This Dirac-K\"{a}hler mechanism
can be understood in the following. In two dimensions
Lorentz generator is represented by one component generator $J$ satisfying 
\begin{align}
 [P_{\mu},J]&=-i\epsilon_{\mu\nu}P_{\nu}\,,\\
[Q_{\alpha i},J]&=-\frac{1}{2}\gamma^{5}_{\alpha\beta}Q_{\beta i}\label{b3}\,.
\end{align}
On the other hand we can rewrite (\ref{a8}) in the same form as
(\ref{b3}) because $S_{ij}$ and $\gamma^{5}_{ij}$ are both
antisymmetric and thus can be chosen to be proportional
to each other,
\begin{equation}
 [Q_{\alpha i},R]=-\frac{1}{2}\gamma^{5}_{i j}Q_{\alpha j}\,.\label{b4}
\end{equation}
Thus we can define $J^{\prime}= J+R$ and obtain 
\begin{equation}
 [s,J^{\prime}]=0\,,\hspace*{10pt}
[s_{\mu},J^{\prime}]=-i\epsilon_{\mu\nu}s_{\nu}\,,\hspace*{10pt}
[\tilde{s},J^{\prime}]=0\,,
\end{equation}
from (\ref{b1}), (\ref{b3}) and (\ref{b4}). 
These relations mean that twisted supercharges $s,s_{\mu}$ and $\tilde{s}$ transform as
scalar, vector and (pseudo-)scalar under $J^{\prime}$, respectively.

As can be seen above the equivalence between Lorentz group and R-symmetry
group is required to realize Dirac-K\"{a}hler twist. 
R-symmetry group is inevitably a compact group and thus 
Lorentz group need to be also compact. There is a natural reason that Euclidean spacetime is
chosen.

The algebra among the twisted supercharges is derived from
(\ref{a7}) and (\ref{b1}) as follows:
\begin{align}
&\{s,s_{\mu}\}=P_{\mu}\,,\hspace*{10pt}
 \{\tilde{s},s_{\mu}\}=-\epsilon_{\mu\nu}P_{\nu}\,,\hspace*{10pt}
 \{s,\tilde{s}\}=0\,,\nonumber\\
&s^{2}=\tilde{s}^{2}=\frac{1}{2}(U_{0}-V_{5})\,,\hspace*{10pt}
 \{s_{\mu},s_{\nu}\}=\delta_{\mu\nu}(U_{0}+V_{5})\,.\label{b10}
\end{align}   
This is the N=2 twisted superalgebra with central charges in two dimensions.

\section{Ansatz on Supercurvature}

In this section we consider so-called superconnection formalism and
find appropriate ansatz on supercurvatures based on the algebra 
derived in the previous section \cite{Saito,KKM}. 

We introduce superfields in the superspace parametrized by $(x_{\mu}, \theta_{A}, z)\,$
\begin{equation}
\Phi(x_{\mu},\theta_{A},z)=\phi(x_{\mu},z)+\theta_{A}\phi_{A}(x_{\mu},z)+\frac{1}{2}\theta_{A}\theta_{B}\phi_{AB}(x_{\mu},z)+\cdots\,,
\end{equation}
where $\theta_{A}$ represents $\theta$, $\theta_{\mu}$ and $\tilde{\theta}$
which are Grassmann coordinates, and $z$ is a real parameter associated
with a central charge.
Using the supercharge differential operator $\mathcal{Q}_{A}$
generating a parameter shift in the superspace, 
we define the supertransformations of the component fields $\phi$, $\phi_{A}$, $\cdots$, 
as follows:
\begin{equation}
\delta_{\xi}\Phi(x,\theta_{A},z)=\delta_{\xi}\phi(x,z)+\theta_{A}\delta_{\xi}\phi_{A}(x,z)+\frac{1}{2}\theta_{A}\theta_{B}\delta_{\xi}\phi_{AB}(x,z)+\dots
\equiv(\xi\mathcal{Q}+\xi^{\mu}\mathcal{Q}_{\mu}+\tilde{\xi}\tilde{\mathcal{Q}})
\Phi(x,\theta_{A},z)\,,
\end{equation}
where $\xi_{A}$ is a Grassmann parameter. 

One can find the supercovariant derivative $\mathcal{D}_{A}$ which anticommutes
with $\mathcal{Q}_{A}$ and then introduce fermionic gauge supercovariant
derivative as
\begin{equation}
 \nabla_{A}=\mathcal{D}_{A}-i\Gamma_{A}\,,\label{c20}
\end{equation}
where $\Gamma_{A}$ is a fermionic superfield called superconnection. 
Similarly bosonic gauged supercovariant derivatives are introduced as
\begin{equation}
\nabla_{\underline{\mu}}=\partial_{\mu}-i\Gamma_{\underline{\mu}}\,,\hspace*{10pt}
\nabla_{z}=\partial_{z}-i\Gamma_{z}\,,\label{c21}
\end{equation}
where $\Gamma_{\underline{\mu}}$ and $\Gamma_{z}$ are bosonic superfields.
The gauge transformation of $\nabla_{I}\equiv\{\nabla_{\underline{\mu}}, \nabla_{A}, \nabla_{z}\}$ is
defined as
\begin{equation}
\nabla_{I}^{\prime}=e^{K}\nabla_{I}e^{-K}\,,\hspace*{10pt} \text{or}\hspace*{10pt}
\delta_{K}\nabla_{I}=[\nabla_{I},K]\,,\label{c30}
\end{equation}
where $K$ is a gauge parameter superfield.
The zeroth order terms of $\Gamma_{A}$ and
$\Gamma_{z}$ w.r.t. $\theta_{A}$ can be taken to be $0$ by choosing Wess-Zumino
gauge while that of $\Gamma_{\underline{\mu}}$ is defined as a usual gauge
connection
\begin{equation}
\Gamma_{A}|=\Gamma_{z}|=0\,,\hspace*{10pt}
\Gamma_{\underline{\mu}}|=A_{\mu}\,,\label{c1}
\end{equation}
where $|$ represents the zeroth order term w.r.t. $\theta_{A}$.
We can thus define standard gauge covariant derivative as 
\begin{equation}
\nabla_{\underline{\mu}}|\equiv
 D_{\mu}=\partial_{\mu}-iA_{\mu}\,.
\end{equation}

We then define the supercurvatures by (anti-)commutation relations of all pair of 
$\nabla_{I}$.
The supercurvatures transform gauge covariantly under (\ref{c30}).
Then some suitable ansatz on supercurvatures leads to an irreducible
supermultiplet. To find such ansatz 
it is useful to introduce supercurvatures $X$, $X^{\prime }$ and 
$X_{\mu}$ defined as
\begin{equation}
\{\nabla_{\alpha i},\nabla_{\beta
 j}\}=-2i\delta_{ij}\gamma^{\mu}_{\alpha\beta}\nabla_{\underline{\mu}}+2\delta_{\alpha\beta}\delta_{ij}X+2\gamma^{5}_{\alpha\beta}\gamma^{5}_{ij}X^{\prime}+2\delta_{\alpha\beta}\gamma^{\mu}_{ij}X_{\mu}\,,\\ \label{c2}
\end{equation}
where $\nabla_{\alpha i}$ are gauged supercovariant derivative
corresponding to $Q_{\alpha i}$ in (\ref{a7}).
Right hand side in (\ref{c2}) is the most general terms for the
consistency with simultaneous replacement of
$\alpha\leftrightarrow\beta$ and $i\leftrightarrow j$.
As can be seen from (\ref{a7}), $X$ and $X^{\prime}$ can be identified as
gauged central charge of $U_{0}$ and $V_{5}$, respectively.
It may be further possible to identify $X$ or $X^{\prime}$ as $\nabla_{z}\,$.
Table \ref{hyou1} shows the relations between gauged supercovariant
derivative and supercurvaturs given in (\ref{c2}) in the twisted space.
It is in principle possible to find different types of supercurvature ansatz.
%One can in fact find different type of ansatz of supercurvatures according to the
%symmetry.
\begin{table}
\centering
\begin{tabular}{c|c|c|c|c|c}
&$\nabla$&$\tilde{\nabla}$&$\nabla_{\nu}$&$\nabla_{\underline{\nu}}$&$\nabla_{z}$ \\ \hline
$\nabla$&$X-X^{\prime}$&$0$&$-i(\nabla_{\underline{\nu}}+iX_{\nu})$&$-iF_{\underline{\nu}}$&$iG$ \\
$\tilde{\nabla}$&&$X-X^{\prime}$&$i\epsilon_{\nu\rho}(\nabla_{\underline{\rho}}-iX_{\nu})$&$-i\tilde{F_{\underline{\nu}}}$&$i\tilde{G}$ \\
$\nabla_{\mu}$&&&$\delta_{\mu\nu}(X+X^{\prime})$&$-iF_{\mu\underline{\nu}}$&$iG_{\mu}$ \\
$\nabla_{\underline{\mu}}$&&&&$-iF_{\underline{\mu}\underline{\nu}}$&$iG_{\underline{\mu}}$ \\
$\nabla_{z}$&&&&&$0$
\end{tabular}
 \caption[]{Twisted version of supercurvature ansatz of (\ref{c2}). For example, $\{\nabla,\nabla\}=X-X^{\prime}$\,,\hspace*{3pt} $[\nabla,
 \nabla_{\underline{\mu}}]=-iF_{\underline{\mu}}$\,.
The positions of $\nabla_{\underline{\mu}}$ reflect those of $P_{\mu}$
 in (\ref{b10}).} \label{hyou1}
\end{table}

We eventually find three types of ansatz.
The first ansatz is shown in Table \ref{hyou2}. 
\begin{table}
 \centering
\begin{tabular}{c|c|c|c|c|c}
&$\nabla$&$\tilde{\nabla}$&$\nabla_{\nu}$&$\nabla_{\underline{\nu}}$&$\nabla_{z}$
 \\ \hline
$\nabla$&$-iW+\nabla_{z}$&0&$-i\nabla_{\underline{\nu}}$&$-iF_{\underline{\nu}}$&$iG$ \\
$\tilde{\nabla}$&&$-iW+\nabla_{z}$&$i\epsilon_{\nu\rho}\nabla_{\underline{\rho}}$&$-i\tilde{F_{\underline{\nu}}}$&$i\tilde{G}$ \\
$\nabla_{\mu}$&&&$\pm\delta_{\mu\nu}(iW+\nabla_{z})$&$-iF_{\mu\underline{\nu}}$&$iG_{\mu}$ \\
$\nabla_{\underline{\mu}}$&&&&$-iF_{\underline{\mu}\underline{\nu}}$&$iG_{\underline{\mu}}$\\
 $\nabla_{z}$&&&&&0
\end{tabular}
 \caption{Supercurvature ansatz of A model. Corresponding to the $\pm$
 sign choice of algebra
 $\{\nabla_{\mu},\nabla_{\nu}\}=\pm\delta_{\mu\nu}(iW+\nabla_{z})$, 
we take $X=\nabla_{z}$ for $+$ and $X^{\prime}=-\nabla_{z}$ for $-$. } \label{hyou2}
\end{table}
Here one of $X$ and  $X^{\prime}$ is identified as $\nabla_{z}$.
We call A model ansatz when bosonic scalar supercurvature ($W$ in the case) is placed in 
diagonal positions. 
It is also possible to include $X_{\mu}$ as supercurvatures.
In this case the bosonic vector superurvatures are placed in off-diagonal
positions, which we call B model ansatz.
One naive candidate for B model ansatz is given in Table \ref{hyou3}. 
\begin{table}
 \centering
\begin{tabular}{c|c|c|c|c|c}
&$\nabla$&$\tilde{\nabla}$&$\nabla_{\nu}$&$\nabla_{\underline{\nu}}$&$\nabla_{z}$
 \\ \hline
$\nabla$&$\nabla_{z}$&0&$-i(\nabla_{\underline{\nu}}+F_{\nu})$&$-iF_{\underline{\nu}}$&$iG$ \\
$\tilde{\nabla}$&&$\nabla_{z}$&$i\epsilon_{\nu\rho}(\nabla_{\underline{\rho}}-F_{\rho})$&$-i\tilde{F_{\underline{\nu}}}$&$i\tilde{G}$ \\
$\nabla_{\mu}$&&&$\pm\delta_{\mu\nu}\nabla_{z}$&$-iF_{\mu\underline{\nu}}$&$iG_{\mu}$ \\
$\nabla_{\underline{\mu}}$&&&&$-iF_{\underline{\mu}\underline{\nu}}$&$iG_{\underline{\mu}}$\\
 $\nabla_{z}$&&&&&0
\end{tabular}
\caption{Naive candidate for B model ansatz} \label{hyou3}
 \end{table}
Jacobi identities, however, lead to $G_{I}=G_{\underline{\mu}}=0$,
 which coincides with a model without central charge.  
On the other hand it is possible to formulate two kinds of B model
 ansatz for one central charge by imposing $U_{0}=\pm V_{5}\,$.
In Table \ref{hyou4} and Table \ref{hyou5} we show the two kinds of
 ansatz which we name B (0,0,Z) model ansatz and B (Z,Z,0) model ansatz, respectively.
\begin{table}
\begin{tabular}{cc}
\begin{minipage}{0.5\hsize}
\centering
\begin{tabular}{c|c|c|c|c|c}
&$\nabla$&$\tilde{\nabla}$&$\nabla_{\nu}$&$\nabla_{\underline{\nu}}$&$\nabla_{z}$
 \\ \hline
$\nabla$&$0$&0&$-i(\nabla_{\underline{\nu}}+F_{\nu})$&$-iF_{\underline{\nu}}$&$iG$ \\
$\tilde{\nabla}$&&$0$&$i\epsilon_{\nu\rho}(\nabla_{\underline{\rho}}-F_{\rho})$&$-i\tilde{F_{\underline{\nu}}}$&$i\tilde{G}$ \\
$\nabla_{\mu}$&&&$\delta_{\mu\nu}\nabla_{z}$&$-iF_{\mu\underline{\nu}}$&$iG_{\mu}$ \\
$\nabla_{\underline{\mu}}$&&&&$-iF_{\underline{\mu}\underline{\nu}}$&$iG_{\underline{\mu}}$\\
 $\nabla_{z}$&&&&&0
\end{tabular}
\caption{Supercurvature ansatz of B (0,0,Z) model} \label{hyou4}
\end{minipage}
\hspace*{2pt}
\begin{minipage}{0.5\hsize}
\centering
\begin{tabular}{c|c|c|c|c|c}
&$\nabla$&$\tilde{\nabla}$&$\nabla_{\nu}$&$\nabla_{\underline{\nu}}$&$\nabla_{z}$
 \\ \hline
$\nabla$&$\nabla_{z}$&0&$-i(\nabla_{\underline{\nu}}+F_{\nu})$&$-iF_{\underline{\nu}}$&$iG$ \\
$\tilde{\nabla}$&&$\nabla_{z}$&$i\epsilon_{\nu\rho}(\nabla_{\underline{\rho}}-F_{\rho})$&$-i\tilde{F_{\underline{\nu}}}$&$i\tilde{G}$ \\
$\nabla_{\mu}$&&&$0$&$-iF_{\mu\underline{\nu}}$&$iG_{\mu}$ \\
$\nabla_{\underline{\mu}}$&&&&$-iF_{\underline{\mu}\underline{\nu}}$&$iG_{\underline{\mu}}$\\
 $\nabla_{z}$&&&&&0
\end{tabular}
 \caption{ Supercurvature ansatz of B (Z,Z,0) model} \label{hyou5}
\end{minipage}
\end{tabular}
 \end{table}

Once suitable ansatz on supercurvatures is obtained, 
a set of relations between the supercurvatures can be derived from Jacobi identities
w.r.t. $\nabla_{I}$,
by which degrees of freedom of the
component fields are reduced. We use a notaion $\nabla W\equiv[\nabla,W]\,$.
For example the component fields in an irreducible supermultiplet can
be defined as 
\begin{equation}
W|=A\,,\hspace*{10pt}\nabla W|=\rho\,,\hspace*{10pt}\cdots\,,\hspace*{10pt}
\text{if}\hspace*{10pt}\nabla W\ne 0\,,
\end{equation}
where $W$ represents a supercurvature while $A$ and $\rho$ are
bosonic and  fermionic component field, respectively. 
The supertransformations and central charge transformations are obtained by 
\begin{equation}
sA=s(W|)\equiv\mathcal{Q} W|=\mathcal{D}W|=\mathcal{D}W|-i[\Gamma,W]| =\nabla W| =\rho\,.
\end{equation} 
The third equality holds at the zeroth order of $\theta_{A}$ while 
the fourth equality holds due to the first relation of (\ref{c1}). 
More complicated supertransformations can be defined by more
sophisticated Jacobi identities.
One thus obtain all supertransformations of each component 
field in an irreducible supermultiplet.

\section{Supermultiplets and Actions}

We derive the supermultiplets and the actions for each model 
with supercurvature ansatz found in the previous section. 
A model, B (0,0,Z) model and B (Z,Z,0) model are considered in subsection
4.1, 4.2 and 4.3, respectively.
Note that N=2 supersymmetric theory in two dimensional Euclidean spacetime 
has generally four bosonic and four fermionic degrees of
freedom at the off-shell level.

\subsection{A model}

We now consider the following algebra:
\begin{align}
\{s,s_{\mu}\}=P_{\mu}\,,\hspace*{10pt}
 &\{\tilde{s},s_{\mu}\}=-\epsilon_{\mu\nu}P_{\nu}\,,\hspace*{10pt}
 \{s,\tilde{s}\}=0\,,\nonumber\\
s^{2}=\tilde{s}^{2}=\frac{1}{2}Z\,,\hspace*{10pt}
 &\{s_{\mu},s_{\nu}\}=\pm\delta_{\mu\nu}Z\,,\hspace*{10pt}
[s_{I},Z]=0\,.\label{d31}
\end{align}
where $+$ represents the case $Z=U_{0}$ and $-$ represents the case
$Z=-V_{5}$ of (\ref{b10}) in the
double sign.
 
The corresponding supercharge differential operators for the superspace
are given by
\begin{align}
\mathcal{Q}&=\frac{\partial}{\partial\theta}+\frac{i}{2}\theta_{\mu}\partial_{\mu}+\frac{i}{2}\theta\partial_{z}\,,\nonumber\\
\mathcal{Q}_{\mu}&=\frac{\partial}{\partial\theta^{\mu}}+\frac{i}{2}\theta\partial_{\mu}-\frac{i}{2}\tilde{\theta}\epsilon_{\mu\nu}\partial_{\nu}\pm\frac{i}{2}\theta_{\mu}\partial_{z}\,,\nonumber\\
\tilde{\mathcal{Q}}&=\frac{\partial}{\partial\tilde{\theta}}-\frac{i}{2}\theta_{\mu}\epsilon_{\mu\nu}\partial_{\nu}+\frac{i}{2}\tilde{\theta}\partial_{z}\,,
\end{align}
where $P_{\mu}$ and $Z$ are represented by $-i\partial_{\mu}$ and
$-i\partial_{z}$, respectively.
The supercovariant derivatives are then found as
\begin{align}
\mathcal{D}&=\frac{\partial}{\partial\theta}-\frac{i}{2}\theta_{\mu}\partial_{\mu}-\frac{i}{2}\theta\partial_{z}\,,\nonumber\\
\mathcal{D}_{\mu}&=\frac{\partial}{\partial\theta^{\mu}}-\frac{i}{2}\theta\partial_{\mu}+\frac{i}{2}\tilde{\theta}\epsilon_{\mu\nu}\partial_{\nu}\mp\frac{i}{2}\theta_{\mu}\partial_{z}\,,\nonumber\\
\tilde{\mathcal{D}}&=\frac{\partial}{\partial\tilde{\theta}}+\frac{i}{2}\theta_{\mu}\epsilon_{\mu\nu}\partial_{\nu}-\frac{i}{2}\tilde{\theta}\partial_{z}\,,
\end{align}
where $\{\mathcal{Q}_{A},\mathcal{D}_{B}\}=0\,$. It should be noted that $\mathcal{D}_{A}$ satisfies
the same algebraic relations as (\ref{d31}) with the identification of
$\mathcal{D}_{A}\rightarrow s_{A}$, while $\mathcal{Q}_{A}$ satisfies the similar relations with
the replacements; $\mathcal{Q}_{A}\rightarrow s_{A}$, $P_{\mu}\rightarrow -P_{\mu}$
and $Z\rightarrow -Z$ in (\ref{d31}).

We now consider supercurvature ansatz in Table \ref{hyou2}.
The following relations can be derived by Jacobi identities:
\begin{align}
&\nabla_{\mu}\nabla W=\epsilon_{\mu\nu}\nabla_{\nu}\tilde{\nabla}W\,,\hspace*{10pt}
F_{\underline{\mu}}=-i\nabla_{\mu}W\,,\hspace*{10pt}
\tilde{F}_{\underline{\mu}}=-i\epsilon_{\mu\nu}\nabla_{\nu}W\,,\nonumber\\
&F_{\mu\underline{\nu}}=\pm i\delta_{\mu\nu}\nabla W \mp i\epsilon_{\mu\nu}\tilde{\nabla}W\,,\hspace*{10pt}
F_{\underline{\mu}\underline{\nu}}=\pm\epsilon_{\mu\nu}\tilde{\nabla}\nabla W+\frac{1}{2}\epsilon_{\mu\nu}\epsilon_{\rho\sigma}\nabla_{\rho}\nabla_{\sigma}W\,,\nonumber\\%\hspace*{10pt}
&G=\nabla W\,,\hspace*{10pt}
\tilde{G}=\tilde{\nabla}W\,,\hspace*{10pt}
G_{\mu}=-\nabla_{\mu}W\,,\hspace*{10pt}
G_{\underline{\mu}}=2i\nabla_{\mu}\nabla W-\nabla_{\underline{\mu}}W\,.
\end{align}
In addition to them we need to impose the following relation:
\begin{equation}
\nabla\tilde{\nabla}W=\frac{1}{2}\epsilon_{\mu\nu}\nabla_{\mu}\nabla_{\nu}W\,.\label{d50}
\end{equation}
This relation is not derived by Jacobi identity but interpreted as a
constraint on supercurvatures to kill the reducibility of representation.

The component fields are then defined as
\begin{equation}
W|=\phi\,,\hspace*{10pt}
\nabla W|=\rho\,,\hspace*{10pt}
\tilde{\nabla}W|=\tilde{\rho}\,,\hspace*{10pt}
\nabla_{\mu}W|=\lambda_{\mu}\,,\hspace*{10pt}
\nabla_{z}W|=D\,,\hspace*{10pt}
G_{\underline{\mu}}|=g_{\mu}\,,
\end{equation}
where $\phi, D$ and $g_{\mu}$ are bosonic fields, and $\rho,
\tilde{\rho}$ and $\lambda_{\mu}$ are fermionic fields.
Table \ref{hyou6} shows the supertransformations of each component field.
\begin{table}
 \centering
\hspace*{-10pt}
\begin{tabular}{c|c|c|c|c}
&$s$&$s_{\mu}$&$\tilde{s}$&$Z$ \\ \hline
 $\phi$&$\rho$&$\lambda_{\mu}$&$\tilde{\rho}$&$D$ \\[2pt]
 $A_{\nu}$&$-i\lambda_{\nu}$&$\pm i\delta_{\mu\nu}\rho\mp i\epsilon_{\mu\nu}\tilde{\rho}$&$-i\epsilon_{\nu\rho}\lambda_{\rho}$&$g_{\nu}$\\[2pt]
$\lambda_{\nu}$&$\frac{i}{2}(g_{\nu}-D_{\nu}\phi)$&$\pm\frac{1}{2}\delta_{\mu\nu}D+\frac{1}{2}F_{\mu\nu}$&$-\frac{i}{2}\partial_{\nu\rho}(g_{\rho}-D_{\rho}\phi)$&$-iD_{\nu}\rho+i\epsilon_{\nu\rho}D_{\rho}\tilde{\rho}-i[\phi,\lambda_{\nu}]$\\[2pt]
 $\rho$&$\frac{1}{2}D$&$-\frac{i}{2}(g_{\mu}+D_{\mu}\phi)$&$\mp\frac{1}{4}\epsilon_{\mu\nu}F_{\mu\nu}$&$\mp
 iD_{\mu}\lambda_{\mu}+i[\phi,\rho]$\\[2pt]
 $\tilde{\rho}$&$\pm\frac{1}{4}\epsilon_{\mu\nu}F_{\mu\nu}$&$\frac{i}{2}\epsilon_{\mu\nu}(g_{\nu}+D_{\nu}\phi)$&$\frac{1}{2}D$&$\mp
 i\epsilon_{\mu\nu}D_{\mu}\lambda_{\nu}+i[\phi,\tilde{\rho}]$\\[2pt]
 $D$&$\mp iD_{\mu}\lambda_{\mu}$&$i\epsilon_{\mu\nu}D_{\nu}\tilde{\rho}-iD_{\mu}\rho$&$\mp
 i\epsilon_{\mu\nu}D_{\mu}\lambda_{\nu}$&$\pm D_{\mu}g_{\mu}\mp
 D_{\mu}D_{\mu}\phi$\\
&&&&$\pm 2i\{\lambda_{\mu},\lambda_{\mu}\}+i[\phi,D]$\\[2pt]
 $g_{\nu}$&$\epsilon_{\nu\rho}D_{\rho}\tilde{\rho}-[\phi,\lambda_{\nu}]$&$\epsilon_{\mu\sigma}\epsilon_{\nu\rho}D_{\rho}\lambda_{\sigma}$&$-\epsilon_{\nu\rho}(D_{\rho}\rho+[\phi,\lambda_{\rho}])$&$\pm
 D_{\rho}F_{\nu\rho}-2\epsilon_{\nu\rho}\{\lambda_{\rho},\tilde{\rho}\}$\\[2pt]
 &&$\mp\delta_{\mu\nu}[\phi,\rho]\pm\epsilon_{\mu\nu}[\phi,\tilde{\rho}]$&&$-2\{\lambda_{\nu},\rho\}+i[\phi,D_{\nu}\phi]$
\end{tabular}
  \caption{Supertransformations of A model} \label{hyou6}
\end{table}
Note that 
$F_{\underline{\mu}\underline{\nu}}|\equiv F_{\mu\nu}=i[D_{\mu},D_{\nu}]\,$ 
is a curvature in usual gauge theory. 
Off-shell closure of the supertransformations up to gauge
transformations is shown with the following constraint on the
component fields:
\begin{equation}
 iD_{\mu}g_{\mu}\mp[\phi,D]-\{\lambda_{\mu},\lambda_{\mu}\}\mp\{\rho,\rho\}\mp\{\tilde{\rho},\tilde{\rho}\}=0\,. \label{d4}
\end{equation}
One can regard this constraint as the same type of constraint 
found for D=N=4 SYM case \cite{SSW}.
  
Because of the constraint the degrees of freedom of $g_{\mu}$ can be
regarded as one. 
Thus the bosonic degrees of freedom at the off-shell
level is four ($\phi$, $A_{\mu}$, $D$, $g_{\mu}$).  
Note that the gauge field $A_{\mu}$ has one bosonic degree of freedom at
the off-shell level.   

For Abelian gauge group the constraint (\ref{d4}) becomes simply as
\begin{equation}
\partial_{\mu}g_{\mu}=0\,,
\end{equation}
which can be  solved as
\begin{equation}
g_{\mu}=\epsilon_{\mu\nu}\partial_{\nu} B\,.
\end{equation}
The degrees of freedom is in fact one.
The explicit form of an action which includes field $B$ is 
\begin{align}
S=\displaystyle\int
 d^{2}x\text{Tr}\Bigl(&\pm\dfrac{1}{2}(\partial_{\mu}\phi)^{2}
-\frac{1}{4}F^{2}_{\mu\nu}-\frac{1}{2}D^{2}\pm\frac{1}{2}(\partial_{\mu}B)^{2}\mp2i\lambda_{\mu}(\partial_{\mu}\rho-\epsilon_{\mu\nu}\partial_{\nu}\tilde{\rho})\nonumber\\
&+e(\frac{1}{2}\phi\epsilon_{\mu\nu}F_{\mu\nu}+2\rho\tilde{\rho}+BD+\epsilon_{\mu\nu}\lambda_{\mu}\lambda_{\nu})\Bigr)\,,
\end{align}
where $e$ is a parameter with mass dimension $1\,$.
In this case invariance of the action and closure of superalgebra are
satisfied without constraints.
It is interesting to note that topological BF term is included in the
action.
For non-Abelian gauge group the constraint cannot be solved locally \cite{solving constraint}.

Finally one can find an action for non-Abelian gauge group as
\begin{align}
S=\displaystyle\int
 d^{2}x\text{Tr}\Bigl(&\pm\dfrac{1}{2}(D_{\mu}\phi)^{2}
-\frac{1}{4}F^{2}_{\mu\nu}-\frac{1}{2}D^{2}\pm\frac{1}{2}g^{2}_{\mu}\mp2i\lambda_{\mu}(D_{\mu}\rho-\epsilon_{\mu\nu}D_{\nu}\tilde{\rho})\nonumber\\
&-i\phi\{\rho,\rho\}-i\phi\{\tilde{\rho},\tilde{\rho}\}\pm i\phi\{\lambda_{\mu},\lambda_{\mu}\}\Bigr)\,. \label{eq201}
 \end{align}
It is worth to mention that this action cannot be derived by superspace.

In this subsection we found a SYM formulation with a constraint.
In the next subsections we can find SYM formulations without constraints.

\subsection{B (0,0,Z) model}

The following algebra is considered: 
\begin{align}
\{s,s_{\mu}\}=P_{\mu}\,,\hspace*{10pt}
 &\{\tilde{s},s_{\mu}\}=-\epsilon_{\mu\nu}P_{\nu}\,,\hspace*{10pt}
 \{s,\tilde{s}\}=0\,,\nonumber\\
s^{2}=\tilde{s}^{2}=0\,,\hspace*{10pt}
 &\{s_{\mu},s_{\nu}\}=\delta_{\mu\nu}Z\,,\hspace*{10pt}
[s_{I},Z]=0\,,\label{d3}
\end{align}
where $Z=2U_{0}=2V_{5}$ in (\ref{b10}).
The corresponding supercharge and supercovariant
derivative differential operators are given by
\begin{align}
\mathcal{Q}=\frac{\partial}{\partial\theta}+\frac{i}{2}\theta_{\mu}\partial_{\mu}\,,\hspace*{10pt}
\mathcal{Q}_{\mu}=\frac{\partial}{\partial\theta^{\mu}}+\frac{i}{2}\theta\partial_{\mu}-\frac{i}{2}\tilde{\theta}\epsilon_{\mu\nu}\partial_{\nu}+\frac{i}{2}\theta_{\mu}\partial_{z}\,,\hspace*{10pt}
\tilde{\mathcal{Q}}=\frac{\partial}{\partial\tilde{\theta}}-\frac{i}{2}\theta_{\mu}\epsilon_{\mu\nu}\partial_{\nu}\,,\nonumber\\
\mathcal{D}=\frac{\partial}{\partial\theta}-\frac{i}{2}\theta_{\mu}\partial_{\mu}\,,\hspace*{10pt}
\mathcal{D}_{\mu}=\frac{\partial}{\partial\theta^{\mu}}-\frac{i}{2}\theta\partial_{\mu}+\frac{i}{2}\tilde{\theta}\epsilon_{\mu\nu}\partial_{\nu}-\frac{i}{2}\theta_{\mu}\partial_{z}\,,\hspace*{10pt}
\tilde{\mathcal{D}}=\frac{\partial}{\partial\tilde{\theta}}+\frac{i}{2}\theta_{\mu}\epsilon_{\mu\nu}\partial_{\nu}\,.
\end{align}
B model (0,0,Z) ansatz is shown in Table 4.
The following relations are derived by Jacobi identities:
\begin{align}
G_{\mu}=0\,,\hspace*{10pt}
\nabla F_{\mu}=\epsilon_{\mu\nu}\tilde{\nabla}&F_{\nu}\,,\hspace*{10pt}
\nabla_{\mu}F_{\nu}+\nabla_{\nu}F_{\mu}=\delta_{\mu\nu}\nabla_{\rho}F_{\rho}\,,\hspace*{10pt}
F_{\underline{\mu}}=-i\nabla F_{\mu}\,,\hspace*{10pt}
\tilde{F}_{\underline{\mu}}=i\tilde{\nabla}F_{\mu}\,,\\
F_{\mu\underline{\nu}}&=-\frac{i}{2}\delta_{\mu\nu}(\nabla_{\rho}F_{\rho}-G)+\frac{i}{2}\epsilon_{\mu\nu}(\epsilon_{\rho\sigma}\nabla_{\rho}F_{\sigma}-\tilde{G})\,,\\
F_{\underline{\mu}\underline{\nu}}&=\nabla_{\mu}\nabla
 F_{\nu}-\nabla_{\nu}\nabla F_{\mu}+i[F_{\mu},F_{\nu}]+\frac{1}{2}\epsilon_{\mu\nu}\nabla\tilde{G}\,,\\
&\hspace*{30pt}\nabla G=\tilde{\nabla}\tilde{G}=\nabla\tilde{G}+\tilde{\nabla}G=0\,,\label{d1}\\
\nabla_{z}F_{\mu}=&\frac{1}{2}(\nabla_{\mu}G-\epsilon_{\mu\nu}\nabla_{\nu}\tilde{G})\,,\hspace*{10pt}
G_{\underline{\mu}}=\frac{i}{2}(\nabla_{\mu}G+\epsilon_{\mu\nu}\nabla_{\nu}\tilde{G})\,.
\end{align}
The component fields are defined as
\begin{equation}
F_{\mu}|=\phi_{\mu}\,,\hspace*{10pt}
\nabla F_{\mu}|=\lambda_{\mu}\,,\hspace*{10pt}
\nabla_{\mu}F_{\nu}|=\frac{1}{2}(\delta_{\mu\nu}\rho+\epsilon_{\mu\nu}\tilde{\rho})\,,\hspace*{10pt}
\nabla_{\mu}\nabla F_{\mu}|=D\,,\label{d40}\hspace*{10pt}
\end{equation}
where $\phi_{\mu}$ and $D$ are bosonic fields and $\rho,
\tilde{\rho}$ and $\lambda_{\mu}$ are fermionic fields.

The supertransformations of each component field are derived
straightforwardly. 
In contrast with the previous model where $G_{A}$ is related to $W$,
$G$ and $\tilde{G}$ should satisfy (\ref{d1}) and seem to be independent from $F_{\mu}$ as far as
Jacobi identities are concerned.
Superalgebra is closed up to gauge transformations without constraints at the off-shell level.  

To obtain the action we introduce linear combination of $s_{\mu}$
\begin{equation}
s_{\pm}\equiv s_{1}\pm is_{2}\,,
\end{equation}
which satisfies
\begin{equation}
s^{2}_{\pm}=0\,,\hspace*{10pt}
\{s_{+},s_{-}\}=2Z\,.
\end{equation}
We define $\lambda_{\pm}\equiv\lambda_{1}\pm i\lambda_{2}$ similarly, 
and introduce the notation
$\nabla^{\pm}_{\underline{\mu}}\equiv\nabla_{\underline{\mu}}\pm F_{\mu}$
 and $D^{\pm}_{\mu}\equiv D_{\mu}\pm \phi_{\mu}$ for convenience.
Then one can derive action by using the nilpotency of $s$, $\tilde{s}$
and  $s_{\pm}$.
In fact we can find  $S_{1}$, $S_{2}$ and $S_{3}$ 
satisfying  $sS_{1}=\tilde{s}S_{1}=s_{+}S_{2}=s_{-}S_{3}=0\,$ where
$S_{1}$, $S_{2}$ and $S_{3}$ are not generally identical. However, in the
case of $\nabla\tilde{G}=\tilde{\nabla}G=0$ together with (\ref{d1}), 
we find 
\begin{align}
S_{1}&=\frac{1}{2}\displaystyle\int d^{2}x
 \text{Tr}s\tilde{s}\tilde{\rho}\rho=S_{0}\,,\\
S_{2}&=\frac{1}{2}\displaystyle\int d^{2}x\text{Tr}s_{+}s_{-}\lambda_{+}\lambda_{-}
=S_{0}+\displaystyle\int d^{2}x\text{Tr}\{-\frac{1}{2}(\epsilon_{\mu\nu}D^{-}_{\mu}\lambda_{\nu}-i(-D^{+}_{\mu}\lambda_{\mu}))(G|+i\tilde{G}|)\}\,,\\
S_{3}&=\frac{1}{2}\displaystyle\int d^{2}x\text{Tr}s_{-}s_{+}\lambda_{-}\lambda_{+}
=S_{0}+\displaystyle\int d^{2}x\text{Tr}\{\frac{1}{2}(\epsilon_{\mu\nu}D^{-}_{\mu}\lambda_{\nu}+i(-D^{+}_{\mu}\lambda_{\mu}))(G|-i\tilde{G}|)\}\,,
\end{align}
where
\begin{equation}
S_{0}=\displaystyle\int
 d^{2}x\text{Tr}\{\frac{1}{2}(D_{\mu}\phi_{\nu})^{2}+\frac{1}{4}F^{2}_{\mu\nu}+\frac{1}{2}D^{2}-i\rho D^{+}_{\mu}\lambda_{\mu}-i\tilde{\rho}\epsilon_{\mu\nu}D^{-}_{\mu}\lambda_{\nu}-\frac{1}{4}[\phi_{\mu},\phi_{\nu}]^{2}\}\,,
\end{equation}
which corresponds to the action without a central charge and to the
twisted versions of the action in \cite{D=2 SYM}.
Here it is important to recognize that 
we can find solutions satisfying 
$\nabla\tilde{G}=\tilde{\nabla}G=0$ and (\ref{d1}),
\begin{equation}
G=a\epsilon_{\mu\nu}\nabla^{-}_{\underline{\mu}}\nabla F_{\nu}\,,\hspace*{10pt}
\tilde{G}=-a\nabla^{+}_{\underline{\mu}}\nabla F_{\mu}\,,
\end{equation}
where $a$ is a parameter with mass dimension $-1$.
Moreover the above choice of $G$ and $\tilde{G}$ makes $S_{2}$ and $S_{3}$ identical
\begin{equation}
S_{2}=S_{3}=S_{0}-ia^{-1}\displaystyle\int d^{2}x\text{Tr}\,G|\tilde{G}|\,.
\end{equation}

We can then find the following action satisfying 
$sS=\tilde{s}S=s_{\mu}S=0$, where the supertransformations are given in Table 7,
\begin{equation}
S=S_{0}-ia^{-1}\displaystyle\int d^{2}x\text{Tr}\,G|\tilde{G}|\,.
\end{equation}
\begin{table}
\centering
\begin{tabular}{c|c|c|c|c}
 &$s$&$s_{\mu}$&$\tilde{s}$&$Z$ \\ \hline
 $\phi_{\nu}$&$\lambda_{\nu}$&$\frac{1}{2}(\delta_{\mu\nu}\rho+\epsilon_{\mu\nu}\tilde{\rho})$&$-\epsilon_{\nu\rho}\lambda_{\rho}$&$\frac{1}{2}(\nabla_{\nu}G|-\epsilon_{\nu\rho}\nabla_{\rho}\tilde{G}|)$  \\[2pt]
 $A_{\nu}$&$-i\lambda_{\nu}$&$-\frac{i}{2}\delta_{\mu\nu}\rho+\frac{i}{2}\epsilon_{\mu\nu}\tilde{\rho}$&$-i\epsilon_{\nu\rho}\lambda_{\rho}$&$\frac{i}{2}(\nabla_{\nu}G|+\epsilon_{\nu\rho}\nabla_{\rho}\tilde{G}|)$ \\
&&$+\frac{i}{2}\delta_{\mu\nu}G|-\frac{i}{2}\epsilon_{\mu\nu}\tilde{G}|$&&\\[2pt]
$\lambda_{\nu}$&$0$&$A_{\mu\nu}$&$0$&$-\frac{i}{2}(D^{-}_{\nu}G|-\epsilon_{\nu\rho}D^{+}_{\rho}\tilde{G}|)$ \\[2pt]
$\rho$&$\frac{i}{2}[D^{+}_{\rho},D^{-}_{\rho}]-D$&$\frac{1}{2}(\nabla_{\mu}G|-\epsilon_{\mu\nu}\nabla_{\nu}\tilde{G}|)$&$\frac{i}{2}\epsilon_{\rho\sigma}[D^{-}_{\rho},D^{-}_{\sigma}]$&$\frac{1}{2}(\nabla_{z}G|-\epsilon_{\rho\sigma}\nabla_{\rho}\nabla_{\sigma}\tilde{G}|)$ \\[2pt]
$\tilde{\rho}$&$-\frac{i}{2}\epsilon_{\rho\sigma}[D^{+}_{\rho},D^{+}_{\sigma}]$&$\frac{1}{2}(\nabla_{\mu}\tilde{G}|+\epsilon_{\mu\nu}\nabla_{\nu}G|)$&$-\frac{i}{2}[D^{+}_{\rho},D^{-}_{\rho}]-D$&$\frac{1}{2}(\nabla_{z}\tilde{G}|+\epsilon_{\rho\sigma}\nabla_{\rho}\nabla_{\sigma}G|)$  \\[2pt]
$D$&$-iD^{+}_{\rho}\lambda_{\rho}$&$\frac{i}{2}(D^{+}_{\mu}\rho-\epsilon_{\mu\nu}D^{-}_{\nu}\tilde{\rho})$&$-i\epsilon_{\rho\sigma}D^{-}_{\rho}\lambda_{\sigma}$&$-\frac{i}{2}(D^{-}_{\rho}\nabla_{\rho}G|+\epsilon_{\rho\sigma}D^{+}_{\rho}\nabla_{\sigma}\tilde{G}|)$ \\
&&$-\frac{i}{2}(D_{\mu}G|-\epsilon_{\mu\nu}D_{\nu}\tilde{G}|)$&&$i\{\rho,G|\}+i\{\tilde{\rho},\tilde{G}|\}$ \\
&&&&$-\frac{i}{2}(\{G|,G|\}+\{\tilde{G}|,\tilde{G}|\})$
  \end{tabular}
 \caption{Supertransformations of B (0,0,Z) model. 
$A_{\mu\nu}=\frac{1}{2}\delta_{\mu\nu}D-\frac{i}{2}(D_{\mu}\phi_{\nu}+D_{\nu}\phi_{\mu}-\delta_{\mu\nu}D_{\rho}\phi_{\rho})+\frac{1}{2}(F_{\mu\nu}-i[\phi_{\mu},\phi_{\nu}])\,.$} \label{hyou7}
\end{table}
In Table \ref{hyou7} the following expressions are used:
\begin{equation}
G|=a\epsilon_{\mu\nu}D^{-}_{\mu}\lambda_{\nu}\,,\hspace*{10pt}
\tilde{G}|=-aD^{+}_{\mu}\lambda_{\mu}\,,\nonumber
\end{equation}
\begin{equation}
\nabla_{\mu}G|=a(-\epsilon_{\mu\nu}\{\lambda_{\nu},\bar{\rho}\}+\frac{1}{2}\{\lambda_{\mu},\tilde{G}|\}+\epsilon_{\rho\sigma}D^{-}_{\rho}A_{\mu\sigma})\,,\nonumber
\end{equation}
\begin{equation}
\nabla_{\mu}\tilde{G}|=a(-\epsilon_{\mu\nu}\{\lambda_{\nu},\bar{\tilde{\rho}}\}-\frac{1}{2}\{\lambda_{\mu},G|\}-D^{+}_{\nu}A_{\mu\nu})\,,\nonumber
\end{equation}
\vspace{-18pt}
\begin{align}
Z\rho&=\frac{a}{2}([\nabla_{\mu}\tilde{G}|,\lambda_{\mu}]+\epsilon_{\mu\nu}[\nabla_{\mu}G|,\lambda_{\nu}]-2[D,\bar{\tilde{\rho}}]+\epsilon_{\mu\nu}[A_{\mu\nu},G|]\nonumber\\
&-i\epsilon_{\mu\nu}D^{+}_{\mu}D^{+}_{\nu}\bar{\rho}-iD^{+}_{\mu}D^{-}_{\mu}\bar{\tilde{\rho}}-\frac{i}{2}\epsilon_{\mu\nu}D^{-}_{\mu}D^{-}_{\nu}G|-\frac{i}{2}D^{-}_{\mu}D^{+}_{\mu}\tilde{G}|) \,,\nonumber\\
Z\tilde{\rho}&=\frac{a}{2}(-[\nabla_{\mu}G|,\lambda_{\mu}]+\epsilon_{\mu\nu}[\nabla_{\mu}\tilde{G}|,\lambda_{\nu}]+2[D,\bar{\rho}]+\epsilon_{\mu\nu}[A_{\mu\nu},\tilde{G}|]\nonumber\\
&+iD^{-}_{\mu}D^{+}_{\mu}\bar{\rho}-i\epsilon_{\mu\nu}D^{-}_{\mu}D^{-}_{\nu}\bar{\tilde{\rho}}-\frac{i}{2}D^{-}_{\mu}D^{+}_{\mu}G|-\frac{i}{2}\epsilon_{\mu\nu}D^{+}_{\mu}D^{+}_{\nu}\tilde{G}|) \,,
\end{align}
where $\bar{\rho}\equiv\rho-\frac{1}{2}G|$ and
$\bar{\tilde{\rho}}\equiv\tilde{\rho}-\frac{1}{2}\tilde{G}|\,$.

\subsection{B (Z,Z,0) model }

We consider the following algebra: 
\begin{align}
\{s,s_{\mu}\}=P_{\mu}\,,\hspace*{10pt}
 &\{\tilde{s},s_{\mu}\}=-\epsilon_{\mu\nu}P_{\nu}\,,\hspace*{10pt}
 \{s,\tilde{s}\}=0\,,\nonumber\\
s^{2}=\tilde{s}^{2}=Z\,,\hspace*{10pt}
 &\{s_{\mu},s_{\nu}\}=0\,,\hspace*{10pt}
[s_{I},Z]=0\,,
\end{align}
where $Z=2U_{0}=-2V_{5}$. The model is completely similar in
construction to the previous model, we thus show mainly results.
B (Z,Z,0) supercurvature ansatz is shown in Table 5.
The component fields are defined as in (\ref{d40}). 

The key relation derived by Jacobi identity is 
\begin{equation}
\nabla_{\mu}G_{\nu}+\nabla_{\nu}G_{\mu}=0\,.\label{d21}
\end{equation}
Similar to the previous model $G_{\mu}$ is not directly related to $F_{\mu}$ by Jacobi identity, 
it is then necessary to solve (\ref{d21}).
As far as the above relation holds one can derive the supertransformations
of each component field and show off-shell closure up to 
gauge transformations without constraints.

Eventually one can find that the following $G_{\mu}$ satisfy $\nabla_{\mu}G_{\nu}=0\,$,
\begin{equation}
G_{\mu}=a\epsilon_{\rho\sigma}(\nabla_{\underline{\rho}}\nabla_{\sigma}F_{\mu}+[F_{\rho},\nabla_{\mu}F_{\sigma}])\,.
\end{equation}
 The following action has full supersymmetry for the
 supertransformations given in Table \ref{hyou8}:
\begin{equation}
S=S_{0}-ia^{-1}\displaystyle\int d^{2}x\text{Tr}\,\frac{1}{2}\epsilon_{\mu\nu}G_{\mu}|G_{\nu}|\,.
\end{equation}
\begin{table}
\hspace*{-25pt}
\centering
\begin{tabular}{c|c|c|c|c}
 &$s$&$s_{\mu}$&$\tilde{s}$&$Z$ \\ \hline
 $\phi_{\nu}$&$\lambda_{\nu}$&$\frac{1}{2}(\delta_{\mu\nu}\rho+\epsilon_{\mu\nu}\tilde{\rho})$&$-\epsilon_{\nu\rho}\lambda_{\rho}$&$\frac{1}{2}(\nabla G_{\nu}|-\epsilon_{\nu\rho}\tilde{\nabla}G_{\rho}|)$ \\[2pt]
 $A_{\nu}$&$-i\lambda_{\nu}+\frac{i}{2}G_{\nu}|$&$\frac{i}{2}(-\delta_{\mu\nu}\rho+\epsilon_{\mu\nu}\tilde{\rho})$&$-i\epsilon_{\nu\rho}\lambda_{\rho}+\frac{i}{2}\epsilon_{\nu\rho}G_{\rho}|$&$\frac{i}{2}(\nabla G_{\nu}|+\epsilon_{\nu\rho}\tilde{\nabla}G_{\rho}|)$ \\[2pt]
$\lambda_{\nu}$&$\frac{1}{4}(\nabla G_{\nu}|-\epsilon_{\nu\rho}\tilde{\nabla}G_{\rho}|)$&$A_{\mu\nu}$&$\frac{1}{4}(\tilde{\nabla}G_{\mu}|+\epsilon_{\mu\nu}\nabla
 G_{\nu}|)$&$\frac{1}{4}\nabla_{z}G_{\nu}|-\frac{1}{2}\epsilon_{\nu\rho}\nabla\tilde{\nabla}G_{\rho}|$ \\[2pt]
$\rho$&$-iD_{\rho}\phi_{\rho}-D$&$0$&$\frac{i}{2}\epsilon_{\rho\sigma}[D_{\rho},D_{\sigma}]$&$-iD^{-}_{\rho}G_{\rho}|$ \\[2pt]
$\tilde{\rho}$&$-\frac{i}{2}\epsilon_{\rho\sigma}[D^{+}_{\rho},D^{+}_{\sigma}]$&$0$&$iD_{\rho}\phi_{\rho}-D$&$-i\epsilon_{\rho\sigma}D^{+}_{\rho}G_{\sigma}|$ \\[2pt]
$D$&$-iD^{+}_{\rho}\lambda_{\rho}+\frac{i}{2}D_{\rho}G_{\rho}|$&$\frac{i}{2}(D^{+}_{\mu}\rho-\epsilon_{\mu\nu}D^{-}_{\nu}\tilde{\rho})$&$-i\epsilon_{\rho\sigma}D^{-}_{\rho}\lambda_{\sigma}+\frac{i}{2}\epsilon_{\rho\sigma}D_{\rho}G_{\sigma}|$&$\frac{i}{2}(D^{-}_{\rho}\nabla G_{\rho}|+\epsilon_{\rho\sigma}D^{+}_{\rho}\tilde{\nabla}G_{\sigma}|)$ \\
&&&&$-2i\{\lambda_{\rho},G_{\rho}|\}+\frac{i}{2}\{G_{\rho}|,G_{\rho}|\}$
  \end{tabular}
 \caption{Supertransformations of B (Z,Z,0) model. 
$A_{\mu\nu}$ is defined in the same way as that in Table \ref{hyou7}.} \label{hyou8}
\end{table}
In Table \ref{hyou8} the following expressions are used:
\begin{equation}
G_{\mu}|=a(D^{-}_{\mu}\tilde{\rho}+\epsilon_{\mu\nu}D^{+}_{\nu}\rho)\,,\nonumber\hspace*{10pt}
\end{equation}
\begin{equation}
\nabla
 G_{\mu}|=a(-\frac{i}{2}\epsilon_{\rho\sigma}D^{-}_{\mu}[D^{+}_{\rho},D^{+}_{\sigma}]-i\epsilon_{\mu\nu}D^{+}_{\nu}D_{\rho}\phi_{\rho}-\epsilon_{\mu\nu}D^{+}_{\nu}D-2\{\tilde{\rho},\bar{\lambda}_{\mu}\}+\frac{1}{2}\epsilon_{\mu\nu}\{\rho,G_{\nu}|\}) \,,\nonumber
\end{equation}
\begin{equation}
\tilde{\nabla}G_{\mu}|=a(-iD^{+}_{\nu}[D^{-}_{\mu},D^{-}_{\nu}]+iD^{-}_{\mu}D_{\nu}\phi_{\nu}-D^{-}D+2\{\rho,\bar{\lambda}_{\mu}\}+\frac{1}{2}\epsilon_{\mu\nu}\{\tilde{\rho},G_{\nu}|\})\,,\nonumber
\end{equation}
\begin{align}
Z\lambda_{\mu}&=\frac{a}{2}(\frac{1}{2}[\nabla
 G_{\mu}|+\epsilon_{\mu\nu}\tilde{\nabla}G_{\nu}|,\tilde{\rho}]-\frac{1}{2}[\tilde{\nabla}G_{\mu}|-\epsilon_{\mu\nu}\nabla
 G_{\nu}|,\rho]+4\epsilon_{\mu\nu}[K,\bar{\lambda}_{\nu}]\nonumber\\
&-2i\epsilon_{\mu\nu}D^{-}_{\nu}D^{+}_{\rho}\bar{\lambda}_{\rho}+2i\epsilon_{\mu\nu}D^{+}_{\rho}D^{-}_{[\nu}\bar{\lambda}_{\rho]}+\frac{i}{2}\epsilon_{\mu\nu}D^{-}_{[\nu}D^{-}_{\rho]}G_{\rho}|
-\frac{i}{2}\epsilon_{\rho\sigma}D^{-}_{\mu}D^{+}_{\rho}G_{\sigma}|\nonumber\\
&-\frac{i}{2}\epsilon_{\mu\nu}D^{+}_{\nu}D^{-}_{\rho}G_{\rho}|-\frac{i}{2}\epsilon_{\rho\sigma}D^{+}_{\rho}D^{+}_{\sigma}G_{\mu}|)\,,
\end{align}
where [ , ] denotes the antisymmetrization of suffixes and
$\bar{\lambda}_{\mu}\equiv\lambda_{\mu}-\frac{1}{4}G_{\mu}|\,$. 

\section{Conclusion and Discussions}

We have constructed off-shell invariant N=2 twisted SYM theory with a
 gauged central charge in two dimensions. Depending on the
 supercurvature ansatz we have introduced A and B models.

In A model superalgebra is closed at off-shell level with an extra
 constraint (\ref{d4}). This model has a similarity with D=N=4 SYM 
theory with gauged central charge with 
unavoidable extra constraint. For Abelian gauge group we can
 explicitly solve the constraint (\ref{d4}), we can thus construct the
 off-shell supertransformations and action without any other constraints.
It is interesting to note that the action has two dimensional
 topological BF term. We cannot, however, solve the constraint for non-Abelian
 case.

On the other hand we have found two types of B model whose superalgebra
 is closed at the off-shell level without any constraints. This gives us a
 hope that we may use the similar ansatz of B model in four dimensions
 to get off-shell invariant N=4 formulation without constraints \cite{to appear}.

\vspace{1cm}
{\bf{\Large Acknowledgements}}

We would like to thank Y.~Kondo and J.~Saito for fruitful discussions,
in particular for J.~Saito for crucial comments. One of the authors (N.K.)
would like to thank I.~L.~Buchbinder and E.~A.~Ivanov for useful discussions. 
This work was supported in part by Japanese Ministry of Education,
Science, Sports and Culture under the grant number 22540261.

\end{document}